\documentclass[10pt,conference]{IEEEtran}
\IEEEoverridecommandlockouts
\usepackage{cite}
\usepackage{multirow}
\usepackage{graphicx}
\usepackage{rotating}
\usepackage[export]{adjustbox}
\usepackage{amsmath,amssymb,amsfonts}
\usepackage{graphicx}
\usepackage{algorithm}
\usepackage{algpseudocode}
\usepackage{booktabs}
\usepackage{afterpage}
\usepackage{textcomp}
\usepackage{pdfpages}
\usepackage{caption}
\usepackage{subcaption}
\usepackage{xcolor}
\usepackage{placeins}
\usepackage{multicol}
\usepackage{url}
\usepackage{lipsum}  
\usepackage{verbatim}
\usepackage[inline]{enumitem}
\usepackage{setspace}
\usepackage{float}
\usepackage{tabularx}
\usepackage{adjustbox}
\usepackage{array}
\usepackage{enumitem}   
\usepackage{array}
\usepackage{multirow}

\newcolumntype{P}[1]{>{\centering\arraybackslash}p{#1}}
\renewcommand\thesubsection{\Alph{subsection}}

\usepackage[bookmarks=false]{hyperref}
\begin{document}

\title{Enhancing QoE in HTTP/3 using EPS Framework}

\author{\IEEEauthorblockN{\textsuperscript{} Abhinav Gupta and Radim Bartos}
\IEEEauthorblockA{\textit{Department of Computer Science} \\
\textit{University of New Hampshire}\\
Durham, NH 03824, USA \\
{\{ag1226,rbartos\}}@cs.unh.edu}}

\maketitle
\begin{abstract}

HTTP/3, the latest evolution of the Hypertext Transfer Protocol, utilizes QUIC, a new transport protocol leveraging UDP to overcome limitations such as connection time and head-of-line blocking prevalent in HTTP/2. This advancement is enhanced by the Extensible Prioritization Scheme (EPS), which introduces a flexible prioritization framework for improving website resource delivery. This paper proposes a mixed scheduling mechanism that delivers using mixed incremental and non-incremental resource delivery and adheres to EPS urgency levels to improve the QoE. Additionally, we propose an EPS priority mapping to enhance the QoE further. This mapping is based on the priority indicated by the Chromium browser and the resource type. The result of the experimental evaluation indicates that the proposed mechanism and mapping improve commonly-used website performance measures for sites featuring a comparatively large number and size of resources.

\end{abstract}


\section{Introduction}
HTTP/3~\cite{rfc9114} represents the latest evolution in the Hypertext Transfer Protocol series, serving as its third major release. This protocol is engineered to enhance the performance and security of web communications. Unlike its predecessors, HTTP/3 operates over QUIC~\cite{rfc9000}, a transport layer network protocol based on UDP. 

To further enhance HTTP/3’s performance capabilities, the Extensible Prioritization Scheme (EPS)~\cite{rfc9218} was introduced. This scheme provides a more flexible and detailed framework for prioritizing resources. Traditional HTTP/2 prioritization often led to inefficiencies and suboptimal resource delivery~\cite{http2_prioritization_issues}. HTTP/3 and the Extensible Prioritization Scheme offer a robust solution for modern internet usage, emphasizing speed, efficiency, and adaptability.

EPS for HTTP has laid out two data transfer methodologies: \emph{Non-Incremental (N)} or \emph{Incremental (I)}. In Incremental transfer, data is sent in chunks that can be used as soon as received, benefiting resources critical to initial content rendering. Non-incremental transfer, on the other hand, delivers data in a single block. Additionally, EPS introduces a range of urgency levels, from 0 to 7, to prioritize resources. An urgency level of 0 is the highest priority, whereas a level of 7 represents the lowest priority.

We have separately explored the performance using non-incremental~\cite{gupta2024improving2} or incremental~\cite{gupta2024improving} resource delivery in our previous research studies. This paper extends this investigation and makes the following contributions:
\begin{enumerate*}[label=(\roman*)]
  \item a mixed scheduling mechanism that combines both N and I requests.
  \item an EPS priority mapping for various web resource types to improve web performance.
\end{enumerate*}
The effectiveness of the proposed methods is evaluated on eight popular websites using Lighthouse, an industry-standard tool developed by Google~\cite{lighthouse} and utilized in previous studies~\cite{saif2020early,gupta, gupta2024improving}. The proposed methods is compared with sequential, non-incremental prioritized, and incremental prioritized delivery methods.

\begin{figure*}[htbp]
  \centering
  \includegraphics[width=\textwidth]{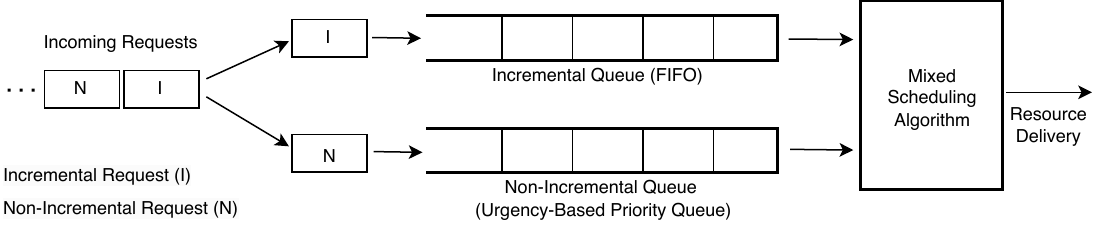}
  \caption{Proposed Mixed Scheduling Mechanism}
  \label{scheduling_mechanism}
\end{figure*}

\begin{figure*}[htbp]
  \centering
  \includegraphics[width=\textwidth]{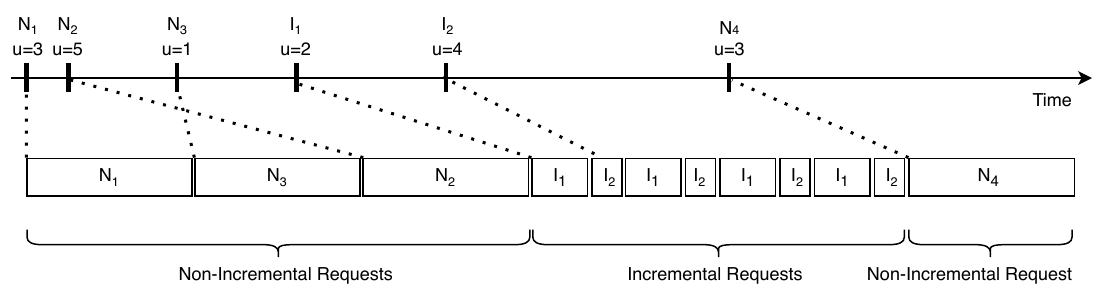}
  \caption{Example of Request Scheduling}
 
  \label{scheduling_example}
\end{figure*}

\section{Related Work}
\label{sec:related_work}
Performance studies~\cite{gupta,dissecting_QUIC,QUICDesignInternetScale,LonglookatQUIC,saif2020early, WebQUICFaster,performance_perspective_quic} of HTTP/3 and QUIC have been conducted extensively following their standardization. The standardization of EPS has redirected focus toward HTTP prioritization, which is used for signaling the urgency and delivery method (N or I) of web requests. 

Wijnants et al.~\cite{h2_wijnants} noted the variability in browser prioritization and the complexity of server-side reprioritization, stressing the need for a more standardized approach in this field. 
Marx et al.~\cite{marx_2019_resource} examined HTTP/3's multiplexing and prioritization, identifying significant performance variations due to web page structures and network conditions. Furthermore, Marx et al.~\cite{same_standards} lists the range of scheduling algorithms these protocols can use.

Sander et al.~\cite{sander2022analyzing} suggest that sequential scheduling is effective in bursty loss scenarios, whereas parallel scheduling enhances performance in situations with increasing random loss. In cases of moderate loss, prioritized parallel scheduling performs better than round-robin. Our previous study demonstrated that \emph{non-incremental prioritized}~\cite{gupta2024improving2} delivery, aligned with the EPS standard, effectively enhances QoE as measured by Lighthouse. Likewise, \emph{incremental prioritized}~\cite{gupta2024improving} delivery study following the same EPS framework also yielded improvements in the QoE.


\section{Scheduling mechanism}
\label{sec:scheduling_mechanism}

In today's era of complex web pages, various HTTP/3 implementations utilize distinct scheduling mechanisms~\cite{same_standards}. The standardization of EPS, as detailed in RFC 9218~\cite{rfc9218}, enables HTTP/3 implementations to prioritize tasks based on the urgency levels of requests and whether the requests are N or I. This EPS framework can improve resource delivery, enhance QoE by reordering requests, and achieve just-in-time delivery.

This paper introduces a mixed scheduling mechanism that utilizes urgencies and manages N and I requests. Grounded in the principles of the Extensible Prioritization Scheme (EPS)  defined in RFC 9218~\cite{rfc9218}, this mixed scheduling mechanism is designed to improve web resource delivery by prioritizing requests based on whether they are N or I and their urgency level. The framework of our proposed mechanism is illustrated in Figure~\ref{scheduling_mechanism}.

There are two distinct queues for requests: the Non-Incremental Priority Queue and the Incremental Queue, each serving different types of web requests. Non-incremental requests, which require complete processing, are prioritized based on urgency and placed in the Non-Incremental Priority Queue. In contrast, incremental requests are added to the Incremental Queue; however, it is crucial to note that the scheduler does the actual reordering of these incremental requests into chunks based on the urgency of the request, not the queue. The incremental request with higher urgency gets a higher share of bandwidth. The mixed scheduling algorithm is shown in Algorithm~\ref{algo_scheduler}.

\begin{algorithm}
\caption{Mixed Scheduling Algorithm}
\label{algo_scheduler}
\begin{algorithmic}[1]
    \While{True} 
        \While{the \textit{Non-Incremental Queue} is not empty}
         \State Deliver \textit{Non-Incremental Priority Queue} requests  
         \Statex\hspace{\algorithmicindent}\hspace{\algorithmicindent}in order of their urgency levels

        \EndWhile

        \While{the \textit{Incremental Queue} is not empty}
            \State Deliver all enqueued requests from the 
            \Statex\hspace{\algorithmicindent}\hspace{\algorithmicindent}\textit{Incremental} \textit{Queue} with bandwidth shares
            \Statex\hspace{\algorithmicindent}\hspace{\algorithmicindent}allocated according to their respective urgency 
\Statex\hspace{\algorithmicindent}\hspace{\algorithmicindent}levels.
        \EndWhile

    \EndWhile
\end{algorithmic}
  \vspace{-2pt}
\end{algorithm}

Figure~\ref{scheduling_example} shows as example of a sequence of N and I requests scheduled using the proposed mechanism. As time progresses, requests are continuously received. Each request carries information about whether it is N or I and its urgency level. Two more N requests follow the first N request. Subsequently, two I requests are received with their corresponding urgencies, and finally, a N request. 

Requests are queued in their respective queues. When the non-incremental queue is not empty, its requests are served; similarly, when the incremental queue is not empty, its requests are processed.


Non-incremental requests are transmitted in their entirety and $N_{2}$ is scheduled after $N_{3}$ due to lower urgency of $N_{2}$ as compared to $N_{3}$. The incremental requests are delivered in an interleaved manner. $I_{1}$ request gets more share of bandwidth as it has higher urgency as compared to $I_{2}$.

\section{Proposed Mapping}
\label{sec:proposed_mapping}
\begin{figure*}[htbp]  
  \centering
  
  \begin{minipage}[b]{0.48\textwidth}  
   \begin{subfigure}[b]{\textwidth}
    \includegraphics[width=\textwidth]{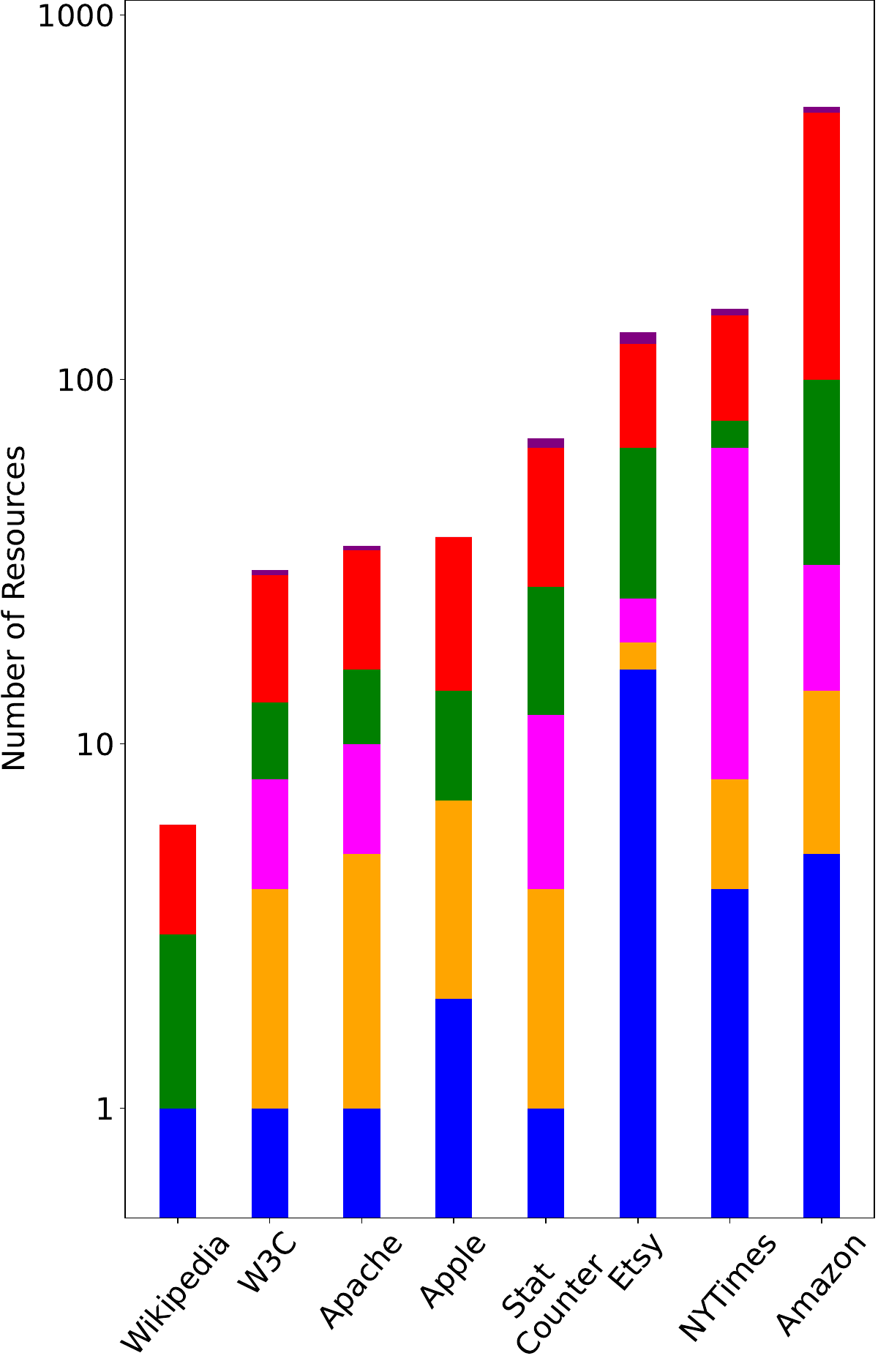}
    \caption{Total Number of Resources }
    \label{websites_sorted_by_count}
     \end{subfigure}
  \end{minipage}%
  \hfill  
  \begin{minipage}[b]{0.48\textwidth}
    \begin{subfigure}[b]{\textwidth}
      \includegraphics[width=\textwidth]{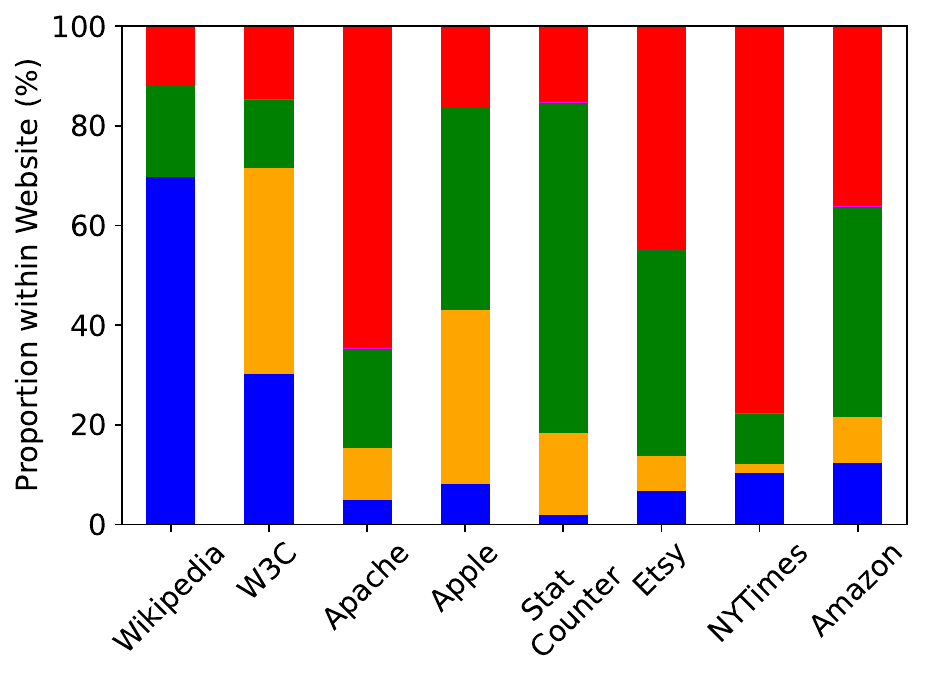}
      \caption{Distribution of Resource Types by Size}
        
      \label{By_size}
    \end{subfigure}
    
    \vspace{10pt}  
    
    \begin{subfigure}[b]{\textwidth}
      
      \includegraphics[width=\textwidth]{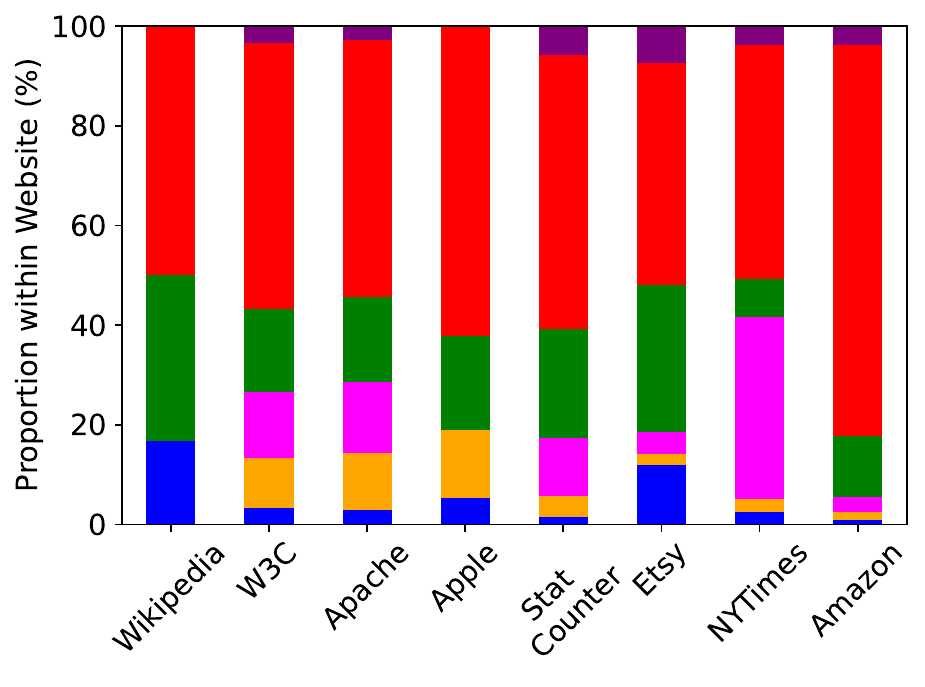}
      \caption{Distribution of Resource Types by Count}
      \label{By_count}
       
    \end{subfigure}
  \end{minipage}
  
  \vspace{10pt} 
  \includegraphics[width=.9\textwidth]{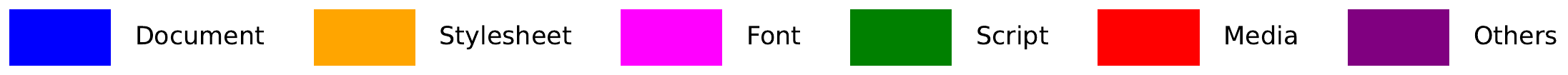} 
   \vspace{10pt} 
  \caption{Websites used in Experimental Evaluation}
  \label{websites_used_in_experiments}
\end{figure*}

The performance of the mixed scheduling mechanism proposed in the previous section is determined by the effective distribution between N and I requests and the appropriate assignment of urgencies. To this end, we examined the priorities assigned by Chromium to resources of eight popular websites, as illustrated in Figure~\ref{websites_sorted_by_count}. Figures~\ref{By_count} and~\ref{By_size} show the eight websites' (wikipedia.org, w3.org, apache.com, apple.com, statcounter.com, etsy.com, nytimes.com, amazon.com) relative distribution by count and size. Chromium assigns Documents and Stylesheets very high priority while not assigning specific priorities to other categories, leaving the levels High, Medium, Low, and Very Low unallocated. If other resources like fonts, scripts, media, and others do not have an assigned urgency value or a specified delivery method N or I at a particular priority level, they have not been given such priority by Chromium and, therefore, are not shown in the resulting EPS Mapping in Table~\ref{Mapping_Table}.

To establish this mapping, we first investigated how different combinations of urgency levels and N or I resource delivery methods could enhance individual Lighthouse metrics such as First Contentful Paint (FCP), Largest Contentful Paint (LCP), Time to Interactive (TTI), Total Blocking Time (TBT), Speed Index (SI), and Cumulative Layout Shift (CLS), as discussed below.

For improving FCP, which measures the time it takes for the first content of the page to be visible, urgency is given to loading the essential structural elements such as documents, stylesheets, and fonts non-incrementally at the highest urgency. This ensures that the DOM (Document Object Model) and CSSOM (CSS Object Model) are quickly constructed, allowing the initial view of the page to render without waiting for all resources. Scripts that affect the initial load are executed incrementally to balance load time and interactivity, while less critical media are also loaded incrementally to prevent them from blocking the early render. Secondly, To improve LCP, which focuses on the loading time of the largest content element visible, the strategy prioritizes non-incremental, high-urgency loading for major content-bearing resources such as critical media and scripts. This approach is crucial for elements like large banners or hero images, which are part of media. Non-incremental loading ensures that these large elements are available and rendered quickly.

Additionally, for TTI and TBT, which measure how quickly a page becomes interactive and the duration of long tasks that block the main thread, respectively, the focus is on prioritizing scripts. Critical scripts are loaded non-incrementally at high urgency to ensure they are available and executable as soon as possible, reducing the time until the page becomes interactive. This minimizes the impact of script execution on the page's responsiveness.

\begin{table}[htbp]
\caption{Proposed Mapping from Chromium Priorities and Resource Types to EPS Urgency Levels and N/I Resource Delivery Method}
\label{Mapping_Table}
\centering
\renewcommand{\arraystretch}{1.5}  
\setlength{\tabcolsep}{3.5pt}       
\begin{tabular}{|l|c|c|c|c|c|c|}
\hline
\multicolumn{1}{|l|}{\parbox{1cm}{\vspace{2pt}\centering \textbf{Chrom.} \\ \textbf{Priority}\vspace{2pt}}} & \multicolumn{6}{c|}{\textbf{EPS Priority}} \\
\cline{2-7}
 & \textbf{Docs.} & \parbox{1cm}{\vspace{4pt}\centering \textbf{Style} \\ \textbf{Sheet}\vspace{4pt}} & \textbf{Font} & \textbf{Script} & \textbf{Media} & \textbf{Others} \\
\hline
\multirow{2}{*}{\parbox{1.2cm}{\vspace{4pt}\textbf{Very} \\ \textbf{High}\vspace{4pt}}} & \multirow{2}{*}{0,N} & \multirow{2}{*}{1,N} & \multirow{2}{*}{1,N} & \multirow{2}{*}{} & \multirow{2}{*}{} & \multirow{2}{*}{} \\
 &  &  &  &  &  &  \\
\hline
\textbf{High} &  &  & 2,N & 2,N & 3,I &  \\
\hline
\textbf{Medium} &  &  &  & 3,N & 4,I &  \\
\hline
\textbf{Low} &  &  &  & 4,I & 5,I & 5,I \\
\hline
\multirow{2}{*}{\parbox{1.2cm}{\vspace{4pt}\textbf{Very} \\ \textbf{Low}\vspace{4pt}}} & \multirow{2}{*}{} & \multirow{2}{*}{} & \multirow{2}{*}{} & \multirow{2}{*}{} & \multirow{2}{*}{} & \multirow{2}{*}{6,I} \\
 &  &  &  &  &  &  \\
\hline
\end{tabular}
\end{table}

Furthermore, for SI, which measures the rapidity of content visibility, documents, stylesheets, and fonts are assigned the highest urgency and are loaded non-incrementally to ensure the foundational structure and styles are applied immediately, facilitating a quick visual display. Scripts in the high urgency category are also loaded non-incrementally to activate crucial interactive features quickly. In contrast, those in the medium category and below are loaded incrementally to balance interactivity with ongoing visual rendering. Media in the high urgency category are loaded incrementally to facilitate a progressive display, enhancing the perceived load speed without entirely blocking the page's usability.

For CLS, which focuses on visual stability to prevent elements from shifting during page load, adopting a similar high urgency for foundational resources like documents and stylesheets ensures that the layout is stable from the outset. Scripts and media that might impact the layout are carefully managed. Essential scripts are loaded non-incrementally to minimize their disruptive potential right after they become available, and media are loaded incrementally when less critical to prevent layout shifts as users interact with the page. This strategic loading aligns to minimize layout shifts while still contributing to a swift and visually complete page rendering. 

Based on these observations and considering the conflicting demands of each performance metric, we propose a mapping scheme shown in Table~\ref{Mapping_Table}. Non-incremental, high-urgency loading for documents, stylesheets, and fonts quickly establishes the page framework, which is essential for FCP and LCP and reduces CLS. A mixed N and I resource delivery approach is applied to scripts and media to ensure critical elements load quickly, enhancing interactivity and perceived completeness. Incremental loading for non-essential media and other resources prevents them from impeding the critical path. 

\section{Experimental Setup}
\label{sec:experimental_setup}

This study establishes an experimental setup consisting of two interconnected virtual machines (VMs), a Client VM and a Server VM, configured within a virtual subnet to form a controlled test environment. These VMs are operated on a MacBook Pro equipped with macOS Sonoma version 14.3.1, an M1 CPU, 64 GB of RAM, and Parallels Desktop 19 for MAC as the virtualization software. Each VM is installed with Ubuntu 22.04.2, utilizes a Linux kernel version 5.15.0-76-generic, and is allocated 16 GB of RAM along with 64 GB of storage. 

Figure~\ref{websites_used_in_experiments} shows the eight websites downloaded and used for the evaluation. The experiments were conducted devoid of internet access to remove any external disruptions. To create a controlled testing environment, external trackers and references were eliminated from several of these websites. This measure prevented downloading additional content and preserved the isolation of our testing setup.

The Client VM runs Lighthouse version 11.2.0 and Chromium version 118.0.5993.70, which is utilized to measure the performance of websites. The Server VM operates two versions of the aioquic server: modified \emph{aioquic} server, which implements the proposed mixed scheduling mechanism and mapping, referred to as the \emph{mixed delivery}, and the standard \emph{aioquic} sequential delivery method, referred to as the \emph{standard delivery}. For simulating network conditions, Netem~\cite{NetEm} was applied to the network interfaces of the VMs. A latency of 10 ms and a packet loss of 0.05\% were injected in each direction, which correspond to typical web resource delivery from a content delivery network.

\begin{figure}[htbp] 
    \centering \includegraphics[width=\linewidth]{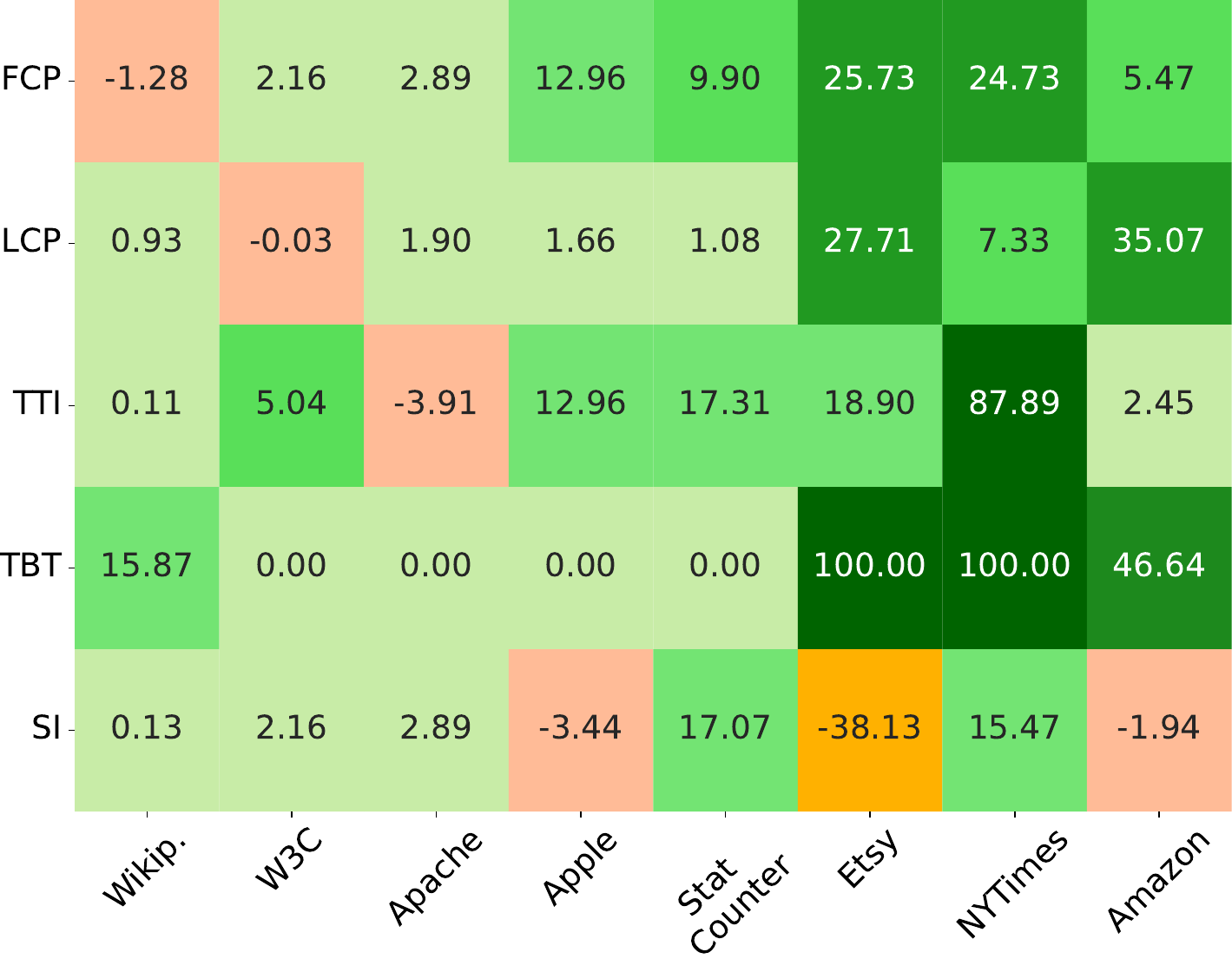} 
    \caption{Relative Performance Improvement of Mixed  over Standard Sequential}
     \label{Mixed_vs_sequential_heatmap}
\end{figure}

\begin{figure}[h!]
  \centering

  \begin{subfigure}[b]{\linewidth}
     \includegraphics[width=\linewidth]{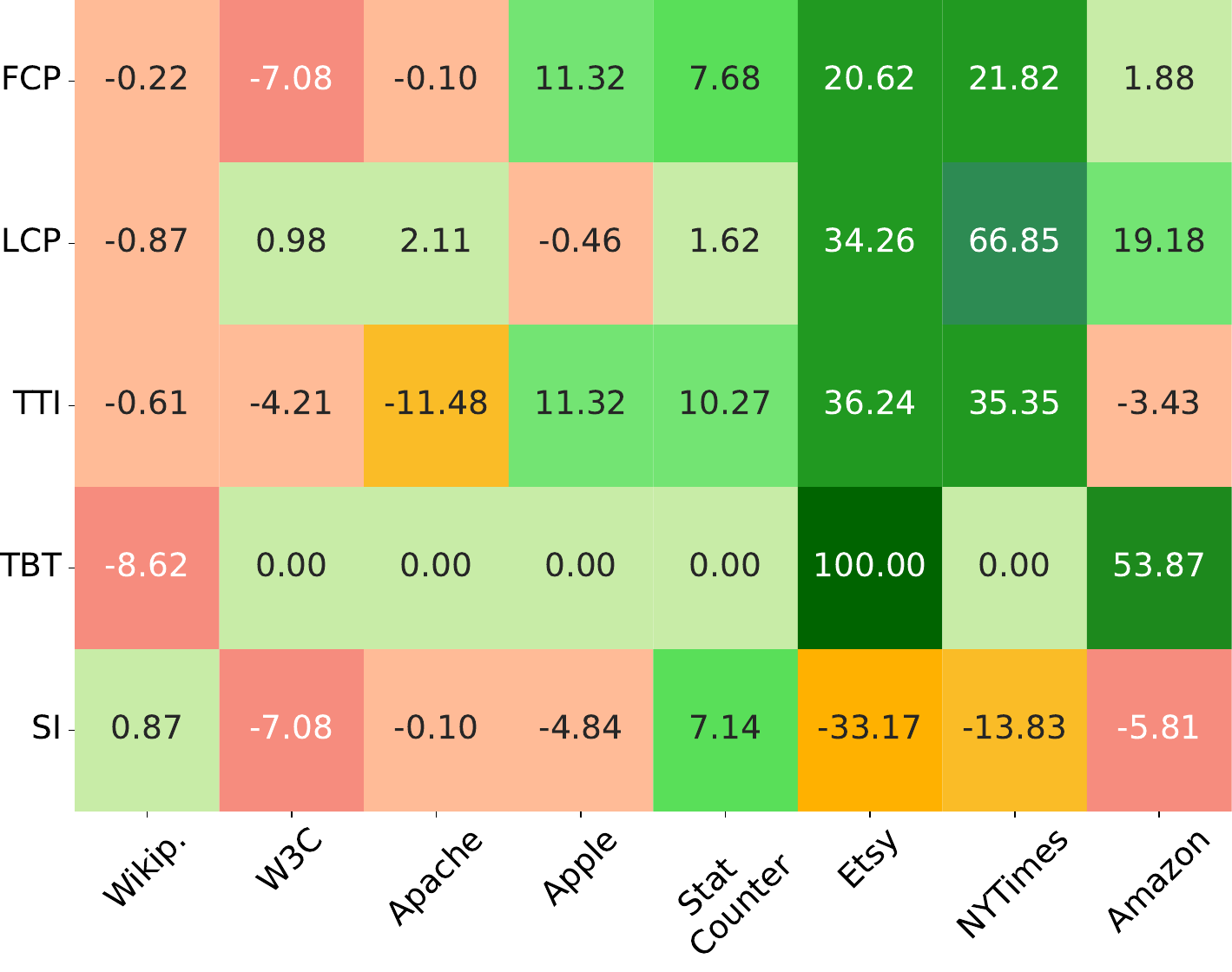}
    \caption{Mixed over Non-Incremental Prioritized }
    \vspace{5pt}
    \label{Mixed_vs_Non-Incremental}
  \end{subfigure}
    \hfill 

  \begin{subfigure}[b]{\linewidth}
   \includegraphics[width=\linewidth]{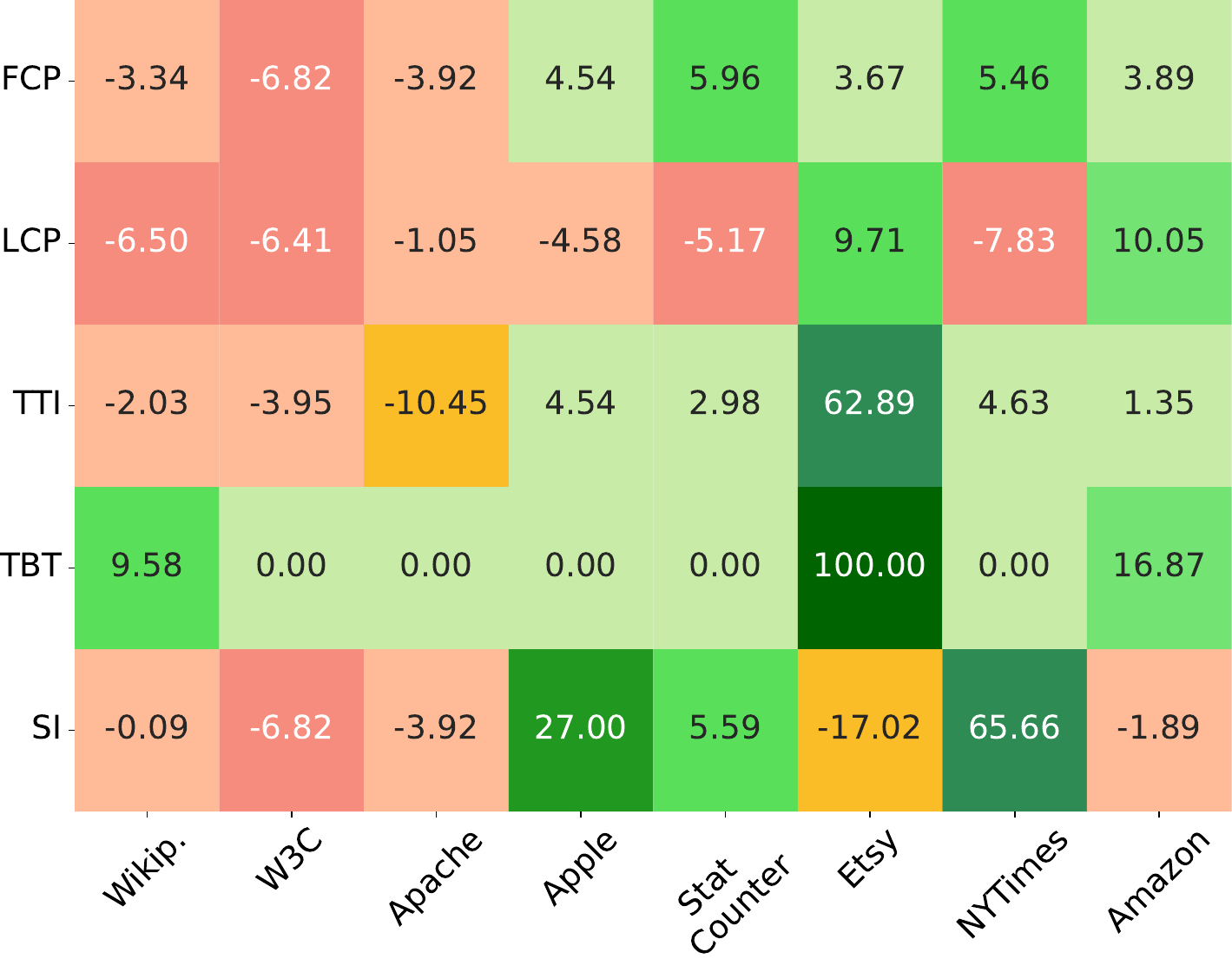}
    \caption{Mixed over Incremental Prioritized}
    \vspace{5pt}
    \label{Mixed_vs_Incremental}
  \end{subfigure}
  \caption{Relative Performance Improvement}
  \label{HeatMap}
  
\end{figure}

\section{Experimental Evaluation}
\label{sec:results}
Figure~\ref{Mixed_vs_sequential_heatmap} illustrates the relative performance improvements across five Lighthouse time based metrics (FCP, LCP, TTI, TBT, and SI) achieved by our mixed delivery compared to the standard delivery. It is to be noted that for time based metrics less time indicates better performance. CLS is not a time-based metric and it is not included in Figure~\ref{Mixed_vs_sequential_heatmap}.
CLS performance will be discussed separately in Subsection~\ref{cls_performance}.

Websites from Wikipedia to Apple are referred to as \emph{small websites} because they have comparatively fewer resources, with Wikipedia having the smallest number. On the other hand, websites from StatCounter to Amazon are termed \emph{large websites} due to their greater number of resources, with Amazon having the most.

\subsection{First Contentful Paint and Largest Contentful Paint}

FCP and LCP are critical components of the Lighthouse performance metrics, contributing significantly to the overall score. Specifically, FCP accounts for 10\%~\cite{lighthouse_performance_scoring1} while LCP contributes 25\%~\cite{lighthouse_performance_scoring1}, together making up 35\% of the total evaluation.

Figure~\ref{Mixed_vs_sequential_heatmap} displays the FCP performance improvements across various websites, showing consistent enhancement from smaller, low-resource sites like Wikipedia to larger sites like Amazon. Etsy and The New York Times exhibit the most significant improvements, each with approximately 25\% enhancement in FCP.

Similarly, the LCP performance also improves as the complexity of the websites increases, with Amazon experiencing the biggest improvement at 35\%.

\subsection{Time To Interactive and Total Blocking Time}

Figure~\ref{Mixed_vs_sequential_heatmap}
shows that the TTI for the mixed delivery performed better than the standard delivery. However, there was a minor decline in Apache performance, highlighting that smaller sites did not substantially improve with mixed delivery. In the most recent update to the Lighthouse performance scoring, TTI has been omitted from the metrics. Despite this change, we have included TTI in our analysis.

TBT, which is heavily weighted in Lighthouse performance scoring, accounting for 30\%~\cite{lighthouse_performance_scoring1} of the total, shows consistent performance improvement. Large websites, specifically Etsy and Amazon, demonstrated a 100\% improvement. Wikipedia, despite being a smaller website, also exhibited a 15.87\% improvement. However, W3C, Apache, Apple, and Stat Counter experienced no change in performance, as TBT was 0 for both the mixed and standard delivery for these websites.

\subsection{Speed Index}

Speed Index accounts for only 10\%~\cite{lighthouse_performance_scoring1} of Lighthouse score. Figure~\ref{Mixed_vs_sequential_heatmap} shows that smaller websites had slightly improved SI performance, but outcomes for larger websites were uneven. Stat Counter and The New York Times showed improvements of 17.07\% and 15.47\%, respectively, whereas Etsy and Amazon did not perform well.

For Etsy, mixed delivery improved FCP, LCP, TTI, and TBT but adversely affected SI. This negative impact is due to complex, dynamic content that, despite faster initial loads, prolonged script processing. Although Amazon's SI slightly deteriorated with the mixed delivery, it remained well below the 3.4 seconds threshold defined as \emph{fast}~\cite{lighthouse_performance_scoring1}, thus not adversely impacting the performance.

\subsection{Cumulative Layout Shift}
\label{cls_performance}
CLS is a unitless metric, accounting for 25\%~\cite{lighthouse_performance_scoring1} of the total Lighthouse score. CLS experienced variability across different websites. Despite the fluctuations, it is essential to note that the CLS scores consistently remained below 0.1, a value considered \emph{good}~\cite{lighthouse_performance_scoring1}, which classify scores up to and including 0.1 as meeting this standard. Consequently, while prioritization of other metrics may have affected the CLS, the visual layout integrity was primarily maintained. Amazon showcased the most significant improvement, with a 73\% improvement in its CLS score.

\subsection{Comparing Mixed with Incremental Prioritized and Non-Incremental Prioritized Delivery}

In order to evaluate the benefits of the mixed delivery, we also present comparison with non-incremental prioritized~\cite{gupta2024improving2} and incremental prioritized~\cite{gupta2024improving} resource delivery. Figure~\ref{Mixed_vs_Non-Incremental} shows the relative performance improvement of mixed over non-incremental prioritized delivery. The results show that the mixed  outperforms non-incremental prioritized delivery, particularly for larger websites. For Etsy, SI performance declines due to prolonged script processing. SI performance of Amazon and the New York Times also fell but remained well within the 3.4-second threshold which is considered as \emph{fast}~\cite{lighthouse_performance_scoring1}.


Similarly, Figure~\ref{Mixed_vs_Incremental} shows the relative performance improvement of mixed over incremental prioritized delivery. It is observed that larger websites benefit more from the mixed delivery method when compared to incremental prioritized delivery. The only marginal decline which we observe is in Apple's LCP score as the incremental prioritized delivery method delivered LCP element quicker. For rest of the websites LCP time remained well below the 2.5-second threshold which is considered \emph{good}~\cite{lighthouse_performance_scoring1}.For Etsy, the decline in SI is attributed to the same issue of prolonged script processing. For SI Amazon experienced some decline but stayed within the 3.4-second threshold which is considered as \emph{fast}~\cite{lighthouse_performance_scoring1}.

\section{Conclusions}
\label{sec:conclusions}

Adopting the Extensible Prioritization Scheme in conjunction with HTTP/3 marks a considerable advancement in web communications, offering an opportunity to improve performance by utilizing the urgency and style of resource scheduling, whether incremental or non-incremental. 

This paper proposes an EPS-based scheduling algorithm that mixes incremental and non-incremental resource delivery based on urgency. The proposed resource mapping further enhances the performance of our scheduling algorithm. Our approach improves performance compared to standard, incremental, and non-incremental HTTP/3 implementations. The most notable improvements are observed in websites with a large number of resources and greater resource size, where our algorithm demonstrates superior responsiveness.

In our future research, we plan to implement the proposed scheduling mechanism in different QUIC implementations. Exploring dynamic adjustment of EPS priorities represents a promising area of investigation.

\bibliographystyle{ieeetr}
\bibliography{bibliography.bib}
\end{document}